\documentclass[%
 reprint,
superscriptaddress,
showpacs,preprintnumbers,
 amsmath,amssymb,
 aps,
]{revtex4-1}

\usepackage{graphicx}
\usepackage{dcolumn}
\usepackage{bm}
\usepackage{hyperref}
\usepackage{float}
\hypersetup{
    colorlinks = true,
    linkcolor=blue,
    citecolor=blue
}
\DeclareMathOperator{\sign}{sign}

\begin{document}

\preprint{APS/123-QED}

\title{Electrically tunable valley polarization in Weyl semimetals with tilted energy dispersion}

\author{Can Yesilyurt}
\email{can--yesilyurt@hotmail.com}
\affiliation{Electrical and Computer Engineering, National University of Singapore, Singapore 117576, Republic of Singapore}

\author{Zhuo Bin Siu}
\affiliation{Electrical and Computer Engineering, National University of Singapore, Singapore 117576, Republic of Singapore}

\author{Seng Ghee Tan}
 \affiliation{Department of Physics, National Taiwan University, Taipei 10617, Taiwan}
\author{Gengchiau Liang}
\affiliation{Electrical and Computer Engineering, National University of Singapore, Singapore 117576, Republic of Singapore}
\author{Shengyuan A. Yang}
 \affiliation{Research Laboratory for Quantum Materials, Singapore University of Technology and Design, Singapore 487372, Singapore}
\author{Mansoor B. A. Jalil}
\email{elembaj@nus.edu.sg}
\affiliation{Electrical and Computer Engineering, National University of Singapore, Singapore 117576, Republic of Singapore}

\date{\today}

\begin{abstract}
Tunneling transport across the p-n-p junction of Weyl semimetal with tilted energy dispersion is investigated. We report that the electrons around different valleys experience opposite direction refractions at the barrier interface when the energy dispersion is tilted along one of the transverse directions. Chirality dependent refractions at the barrier interface polarize the Weyl fermions in angle-space according to their valley index. A real magnetic barrier configuration is used to select allowed transmission angles, which results in electrically controllable and switchable valley polarization. Our findings may pave the way for experimental investigation of valley polarization, as well as valleytronic and electron optic applications in Weyl semimetals.

\end{abstract}

\maketitle

\section{\label{sec:level1}INTRODUCTION}

Charge carriers in crystal lattices may carry a valley isospin degree of freedom, in addition to their real spin. Recent theoretical and experimental works \cite{RN1,RN2,RN3,RN4} have predicted and demonstrated the existence of valley-dependent transport features in condensed matter systems, in which valley polarization has been demonstrated by using magnetic fields \cite{RN5,RN6,RN7,RN8}, line defects \cite{RN9}, and optical helicity \cite{RN10}. A considerable amount of effort has been made to achieve valley dependent tunneling in condensed matter systems. One possible avenue is by applying a uniaxial strain to induce a relative shift of the two valleys at the corners of the hexagonal unit cell of the reciprocal lattice. As a result, the Fermi surfaces experience a shift in  \({k}\)-space, causing the electrons around different valleys to experience refractions in the opposite direction at the interface between the strained and unstrained regions \cite{RN11}. Although  the gradient of uniaxial strain results in the angular separation of electron trajectories according to the valley index, this in itself would not give rise to valley-polarized conductance since the contribution of both valleys to the overall conductance is still identical. To induce a valley-polarized transport, one needs to break the angular symmetry of transmission profile by means of transverse Lorentz displacement \cite{RN12,RN13,RN14,RN15,RN16,RN17}, which may be achieved by applying a magnetic barrier. This forms the basis of various valley transport or filter applications in graphene and silicene \cite{RN12,RN13,RN14,RN15,RN16,RN17}.  However, such an approach does not provide a convenient mechanism to control or modulate the valley transport dynamically. This is because the modulation would entail either a change of the direction of the applied strain or switching of the magnetization direction of the magnetic material. Another way to achieve valley polarized tunneling is the valley-dependent substrate-induced band gap, together with application of gate voltage, and inhomogenous magnetic field profile \cite{RNR4,RNR5}. In this approach, the lifting of valley degeneracy is strongly related to the valley-dependent band structure induced by a particular substrate (e.g., h-BN). Due to the fixed value of the Dirac band gap, this may entail precise tuning of parameters, such as the Fermi energy and applied gate voltage to achieve a finite valley polarization.
Here, we show that valley dependent tunneling can be controlled solely by means of an electrical potential applied to a region of Weyl semimetal with tilted energy dispersion. Theoretically, the proposed valley polarization approach is not restrictive in that it does not require a precise range of parameter values. The mechanisms used to lift the valley degeneracy (i.e., electrical potential barrier) and to select the desired valley transmission (i.e., magnetic barrier) are both tunable. Besides, the previously reported methods for generating valley polarization have been applied to the specific context of graphene. These methods may be material-dependent, e.g. strain, or substrate induced band gap may not be directly applicable to three-dimensional Weyl and Dirac semimetals. Therefore, additionally our work proposes the realization of valleytronic applications in a new material platform, i.e., of Weyl semimetals.

\begin{figure}[b]
\includegraphics[width=85mm]{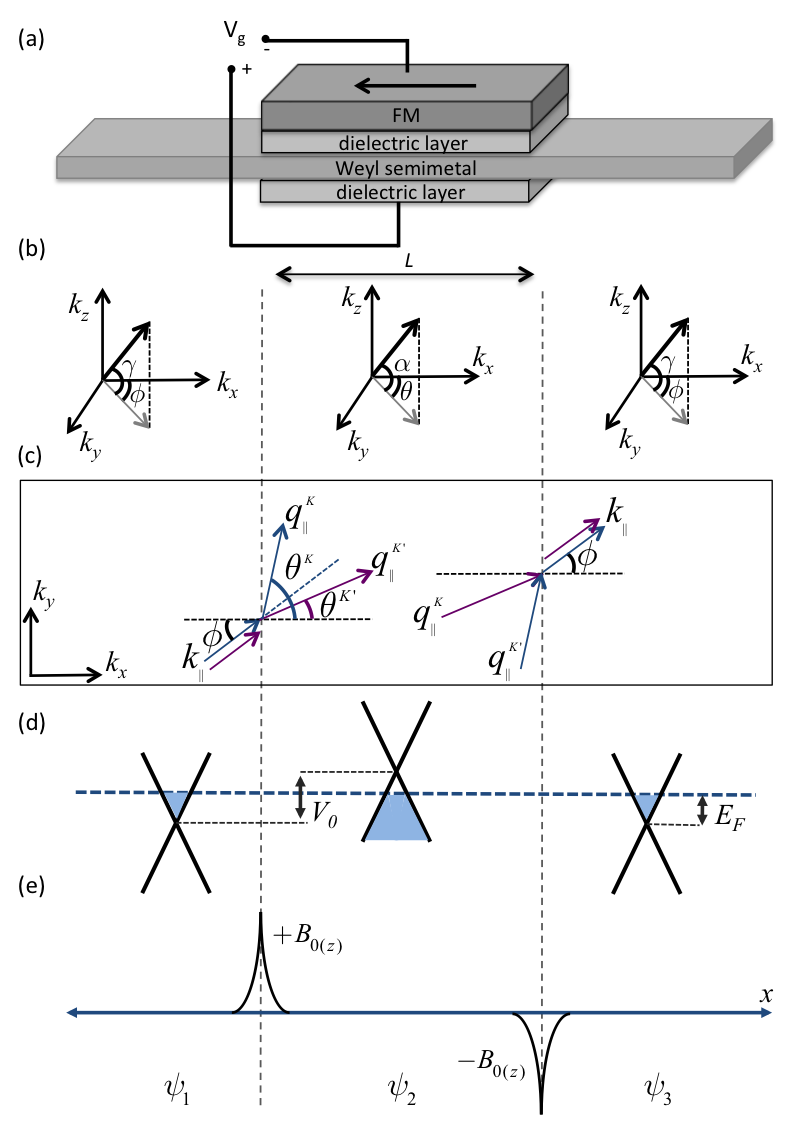}
\caption{\label{fig:epsart} (a) Weyl semimetal with one-dimensional rectangular, square electrical potential and magnetic barrier induced by ferromagnetic (FM) layer deposited on the central region of the p-n-p junction. The black arrow shows the magnetization direction of the FM layer. (b) shows the electron propagation angles outside and within the barrier region. (c) illustrates the valley dependent refractions at the barrier interfaces. (d) and (e) demonstrate the effect of applied electric potential on the central region and magnetic field induced by the FM layer respectively.}
\label{Fig:1}
\end{figure}

It has been previously shown that Weyl fermions that encounter a potential barrier experience an anomalous transverse momentum shift along the direction of the tilt \cite{RN18}. We focus on the simplest Weyl semimetal case where the time reversal symmetry is broken; which allows the presence of a single pair of Weyl nodes emerges related by the inversion symmetry. The valley resolved angular dependence of transmission probability shows that the electrons around different valleys experience deflections along different directions at the barrier interface as illustrated in the Fig. \ref{Fig:1} (c) for an electron incident with momentum  \({k_{||}}\). We consider an additional magnetic barrier configuration applied to the central region of the p-n-p junction to select a particular angle range allowed for transmission, which results in valley-polarized conductance. In this letter, as an example, the magnetic barrier is generated by a FM layer on the top of the barrier region as illustrated in Fig. \ref{Fig:1} (a). While the role of the magnetic barrier is to extract a particular valley polarization, it plays no part in tuning the valley polarization (the magnetic barrier strength is kept constant). We will show that modulation of the valley polarization can be achieved by tuning the applied gate voltage, i.e. by changing the voltage, the valley isospin that contributes to the conduction can be switched.

\section{\label{sec:level1}MODEL}
The electronic states in Weyl semimetals consist of two different characteristics, i.e., Weyl nodes separated in \({k}\)-space and Fermi arc states connecting the projection of two Weyl nodes on a surface \cite{RN19,RN20,RN21,RN22,RN23,RN24}. The contribution of the Fermi arc states to the tunneling conductance is negligible as shown in a Dirac semimetal Na\({_3}\)Bi \cite{RNR1}, and total transmission of the system can be calculated by taking into account only Weyl nodes as considered in various previous works \cite{RN25,RN26,RN27,RN28,RN29}. The robustness of the Weyl semimetal case strongly depends on the \textit{k}-space distance of the Weyl nodes, i.e., the length of the Fermi arc. It has been shown that the Fermi arc length can be tunable \cite{RNR16} and specific materials (e.g. TA\({_3}\)S\({_2}\)) having robust widely separated Weyl nodes have been predicted \cite{RNR17}. In terms of electron transport, these developments would lead to low intervalley scattering effects as long as transmission direction is chosen properly (i.e., discussed in the following sections). Weyl fermion comprises of two linear bands disperse along three dimensions, which are degenerate at a Weyl point. Since Weyl nodes usually occur at generic \({k}\)-points in the Brillouin zone with lower symmetry, Weyl fermions generally possess tilted energy dispersion. A Weyl fermion can be described by a general low-energy Hamiltonian such as

\begin{equation}
\
 H = {V_0} + \sum\limits_i \hbar  {k_i}\tau \left( {{v_i}{\sigma ^i} + {w_i}} \right),
\
\label{Eq:1}
\end{equation}

\noindent
where  \(\sigma \)'s are Pauli matrixes, and \({V_0}\) is external electrical potential. In general, the velocities  \({v}\)'s may be asymmetric in three-dimension, and their sign (\(\tau  =  \pm \)) carry the chirality of Weyl nodes. However, we assume symmetric velocities equal to \({v_F} = {10^6}\)\(m/s\) in the rest of the manuscript. The dispersion of Weyl fermion can be tilted along all three directions, and the strength of the tilt is denoted by \({w_i}\). Based on the proposed approach, the tilt direction must be aligned along one of the transverse directions (\({y}\)-direction in this letter). Therefore, the tilt vector is defined as \(w =(0,\chi w_y, 0 )\), where  \(\chi  =  \pm \) according to the valley index assuming the case where the energy dispersion is tilted along opposite directions in different valleys. We consider ballistic tunneling transmission along the \({x}\)-direction and the incident angles are characterized by \(\gamma \) (the angle between \({k}\) and the \({x}\)-\({y}\) plane) and \(\phi \) (the azimuthal angle with respect to the \({x}\)-axis), as shown in Fig. \ref{Fig:1} (b). Two inequivalent valleys separated in \({k}\)-space are represented by \(K\) and \(K'\). The wave vectors are described by

\begin{equation}\label{eq:2}
\begin{array}{ccl}
{k_x} = {k_F}\cos \gamma \cos \phi , \\
{k_y} = {k_F}\cos \gamma \sin \phi, \\
{k_z} = {k_F}\sin \gamma, 
\end{array}
\end{equation}

where the Fermi wave-vector

\begin{equation}
\
 {k_F} = \left( {{E_F} - {V_0}} \right)/\hbar ({v_F} + {\chi w_y}\cos \gamma \sin \phi ).
\
\label{Eq:3}
\end{equation}

We consider a one-dimensional rectangular potential barrier, which is described by \({V_{(x)}} = {V_0}[\Theta (x) - \Theta (x - L)]\), as illustrated in Fig. \ref{Fig:1} (d). From the practical point of view, such a potential barrier can be induced in Weyl semimetals either by changing the carrier concentration locally by means of electrical gates, and/or doping with alkali metal atoms \cite{RN30,RN31}. Electrostatic gates have been commonly used to tune carrier concentration in two-dimensional materials. Recent experimental works show that dynamic tuning of the carrier concentration is achievable in three-dimensional materials as shown in Dirac semimetal Cd3As2 \cite{RN30} and Weyl semimetal WTe2 \cite{RN32}. The electrostatic gating requires very thin material structure, whose thickness is restricted by the range of the screening effect. Another requirement is that the sample must reach sufficient thickness to allow formation of Weyl nodes in the bulk. For instance, gate bias tuning of the carrier concentration of bulk states has been demonstrated in 50-nm thick Cd3As2 \cite{RN30} and 14-nm thick WTe2 \cite{RN32}.

The FM layer placed on the central region induces two spike-like magnetic fringe fields at the barrier boundaries, as illustrated in Fig. \ref{Fig:1} (e).  Note that a thin film dielectric layer is deposited between the Weyl semimetal and FM layer to avoid induced magnetization in the former due to the proximity effect. To derive an analytical solution of the problem, we first assume the simplified magnetic barrier \cite{RN12,RN13,RN14,RN15,RN33,RN34,RN35} which can be approximately described by two asymmetric delta magnetic fields \({B_{z(x)}} = {B_0}[\delta (x) - \delta (x - L)]\) along the \textit{z}-direction. Later, we will discuss a more realistic magnetic field profile, which shows that the shape of the magnetic field profile does not play a vital role in the proposed method. By the choice of the Landau gauge, the delta magnetic fields give rise to a piecewise constant magnetic vector (gauge) potential \({\vec A_B} = {B_0}{l_B}[\Theta (x) - \Theta (x - L)]\hat y\). The transverse wave vector experience a shift such that \({k_y} \to {k_y} + e{A_B}/\hbar \) within the barrier region. The same magnetic barrier structure can be achieved by various methods and configurations such as two FM strips with asymmetric perpendicular magnetic anisotropy deposited on the barrier boundaries \cite{RN36,RN37,RN38,RN39,RN40,RN41,RN42}, as well as superconductor film fabricated with the desired pattern on the Weyl semimetal and applying uniform magnetic field \cite{RN39,RN43,RN44}. By considering both electrical and magnetic barriers, the momentum along the transmission direction within the barrier is given by

\begin{widetext}
\begin{equation}
\
q_x = \frac{{\sqrt { - {\hbar ^2}v_F^2\left( {{{\left( {{k_y} + \delta {k_y}} \right)}^2} + k_z^2} \right) + {{\left( { - {E_F} + {V_0} + \hbar {\chi w_y}\left( {{k_y} + \delta {k_y}} \right)} \right)}^2}} }}{{\hbar {v_F}}},
\
\label{Eq:4}
\end{equation}
\end{widetext}

\noindent
where \(\delta {k_y} = e{A_B}/\hbar \). Due to the conservation of energy and transverse momentum, electrons experience refractions at the barrier interface, and the propagation angles of the transmitted electrons are given by \(\theta  = {\tan ^{ - 1}}\left( {\frac{{{k_y}}}{{{q_x}}}} \right)\)and \(\alpha  = {\tan ^{ - 1}}\left( {\frac{{{k_z}}}{{{q_x}}}\cos \theta } \right)\) within the barrier region, as illustrated in Fig. \ref{Fig:1} (b). Solving the Hamiltonian in Eq. \ref{Eq:1}, the wave functions of incident, propagated and transmitted electrons are found as

\begin{equation}\label{eq:Eq:5}
{\psi _ \pm } \equiv \frac{1}{{\sqrt 2 }}{e^{i\vec k\vec r}}\left( {\begin{array}{*{20}{c}}1\\{{e^{i\phi }}\sec {\gamma ^{}}\left( {\eta  + \sin \gamma } \right)}\end{array}} \right) \equiv \left( {\begin{array}{*{20}{c}}{{\psi _a}}\\{\psi _b^\eta }\end{array}} \right)
\end{equation}
	
	  The wave function is composed of two spinor components \(\psi _a\)  and \(\psi _b^\eta\) , where \(\eta  = \sign\)\(\left( {{E_F} - {V_0}} \right)\) is the band index. To analyze the angular dependence of electron tunneling we calculate the transmission probability across the system by matching the top and bottom components of the wave functions at the barrier interfaces.  \(T_{(\phi ,\gamma)}^{K}\) and  \(T_{(\phi ,\gamma)}^{K'}\) denote the transmission probability of electrons around  \(K\) and  \(K'\) respectively. Transmission between  \(K \)  and \(\ K' \) is neglected based on the analysis presented in the sections below. We assume equilibrium system, where the source and drain are at the same chemical potential, and thus the Fermi level assumed to be level throughout. To derive the analytical solution, we have neglected the scattering mechanisms in our calculation. However, it has been shown that the angular refractions of electrons (which is a key component of our proposal) and the resultant transmission profile in tilted Weyl systems are robust against weak disorders \cite{RN18}. The fermi surface of the tilted Weyl fermion is not perfect sphere, which must be taken into account in the conductance integral. The infinitesimal element of the elliptical Fermi surface per unit variation of the coordinates \(\phi \) and \(\gamma \) is found as 

\begin{equation}\label{eq:Eq:6}
d{S_{FS}} = \frac{{E_F^2\cos \gamma \sqrt {v_F^2 + (\chi w_y)^2 + 2{v_F}{\chi w_y}\cos \gamma \sin \phi } }}{{{\hbar ^2}{{\left( {{v_F} + {\chi w_y}\cos \gamma \sin \phi } \right)}^3}}}d\phi d\gamma.
\end{equation}

The total valley dependent ballistic conductance is given by Landauer-B{\"u}ttiker formula (detailed derivations are given in Appendix A),

\begin{equation}\label{eq:7}
G_{K(K')}=G_0 \int_{-\frac{\pi }{2}}^{\frac{\pi }{2}}  \int_{-\frac{\pi }{2}}^{\frac{\pi }{2}} d\phi d\gamma   \frac{\cos^2(\gamma) \cos(\phi) }{(1+ \frac{\chi w_y}{v_F} \cos(\gamma) \cos(\phi) )^3  }  T_{(\phi, \gamma)}^{K(K')}
\end{equation}

\noindent
where  \( G_0= \frac{e^2 E_F^2 A}{(2\pi)^3 \hbar^3 v_F^2} \)  is the quantum conductance, and the integral is dimensionless. A is the cross-sectional area of the system. Finally, we define the valley polarization (e.g., for valley \(K\)) \({P_K} = \left( {{G_K} - {G_{K'}}} \right)/\left( {{G_K} + {G_{K'}}} \right)\).

\begin{figure}[b]
\includegraphics[width=84mm]{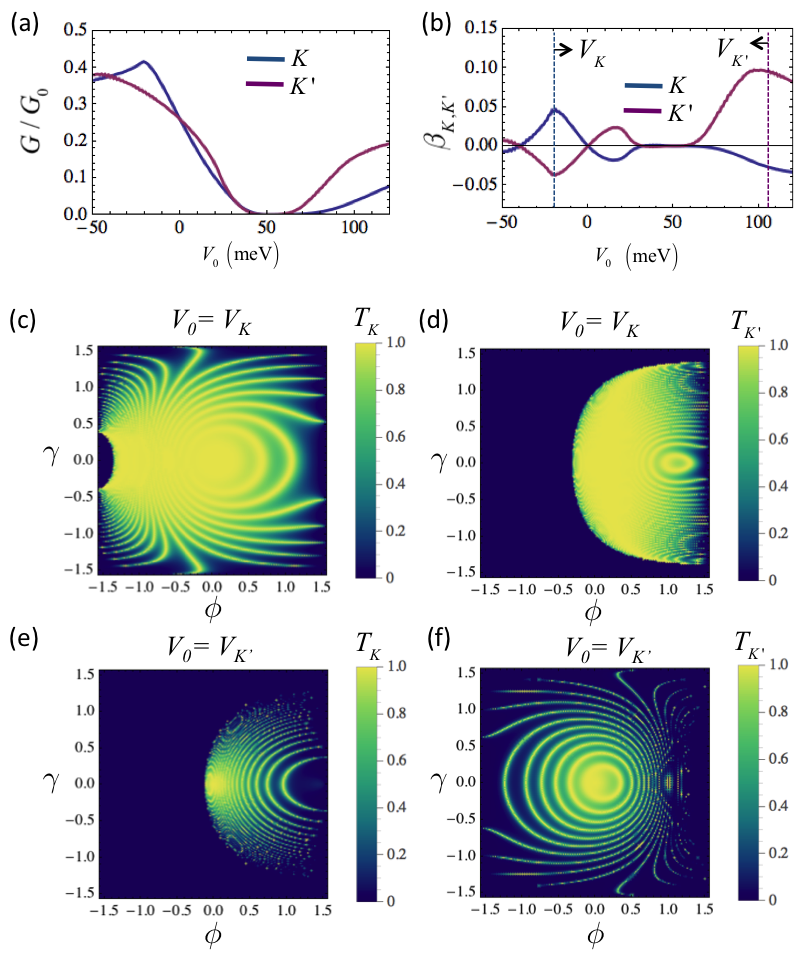}
\caption{\label{fig:epsart} Valley resolved tunneling profile of Weyl electrons around two distinct valleys \(K\) and \(K'\). The normalized tunneling conductance of \(K\) and \(K'\) is shown in (a), where the valleys exhibit different tunneling profiles in the case of varying external electrical potential \(V_0\). The difference between the conductance profile of  \(K\) and \(K'\) gives rise to valley-polarized conductance whose magnitude can be defined by \(\beta_{K,K'}\) as shown in (b). At two chosen external potentials \(V_K\) and  \(V_K'\) [denoted by doted lines in (d)] which yield a high effective polarization   \(\beta_{K,K'}\)  for the  \textit{K} and  \textit{K'} valley, respectively, the angular dependence of tunneling probability for both valleys is shown in (c) to (f). The tilt velocity \(w_y/v_F=\chi \)0.4 , the Fermi energy \(E_F\) =50 meV, barrier length \textit{L}=900 nm, magnetic field  \( B_{z(x)} \)=2 T , and the conductance is in unit of  \(G_0\)  for all configurations.}
\label{Fig:2}
\end{figure}

\section{\label{sec:level1}RESULTS AND DISCUSSIONS}

\begin{figure}[b]
\includegraphics[width=80mm]{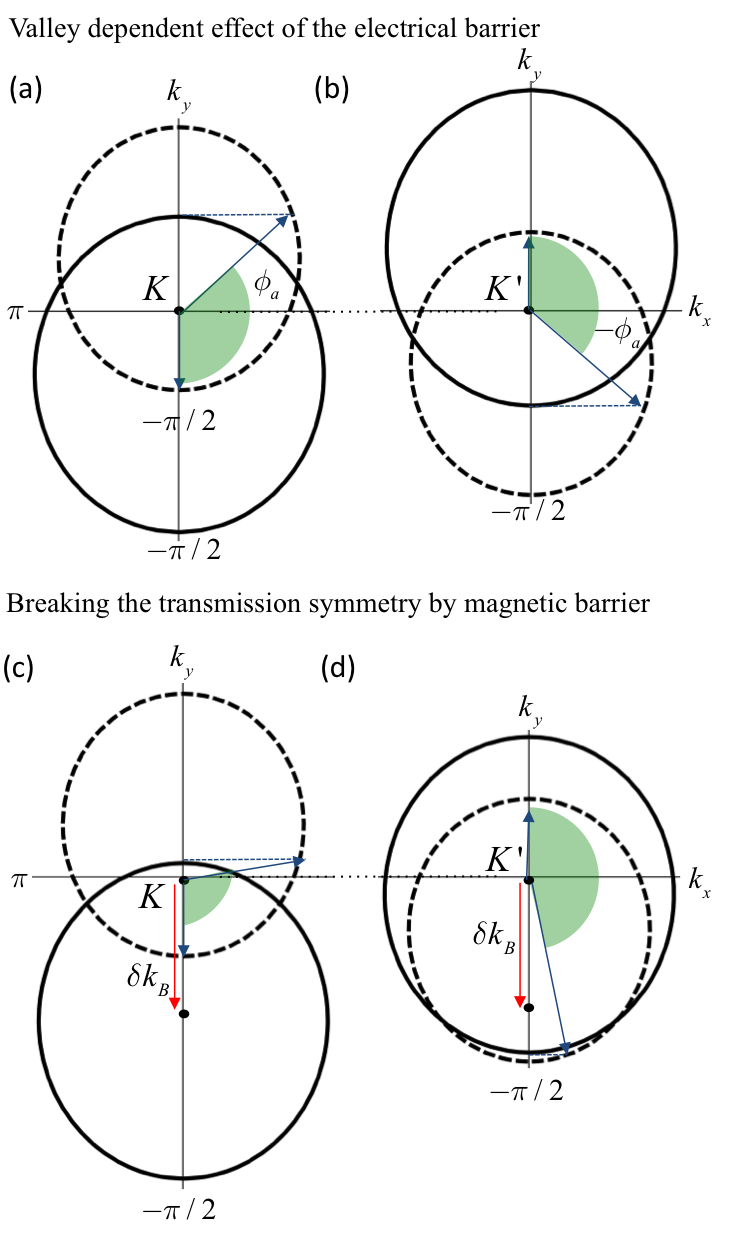}
\caption{\label{fig:epsart} The solid-lined and dashed circles represent the Fermi surfaces of the first and second regions shown in Fig. \ref{Fig:1} (a) respectively. The conservation of transverse wave vector \({k_y}\) limits the allowed transmission range in angle-space in the case of external electrical and magnetic potentials. \(K\) and \(K'\) are two inequivalent valleys which possess opposite chirality and opposite direction tilt (\({w_y}/{v_F} = \chi \)0.4) along the \({y}\)-axis. (a) and (b) illustrate that the applied electrical potential (\({V_0} = 110\) meV) gives rise to a valley dependent momentum shift, causing the valley dependent refraction and reflections at the barrier interface. (c) and (d) demonstrate the further effect of external delta-magnetic field (\({B_0}_{(z)} = 2\) T) induced by FM layer on the central region, causing the contraction (at \(K\)) and extension to all possible angles in the forward direction (at \(K'\)) of the allowed transmission range (denoted by the shaded angles). The Fermi energy \({E_F} = 50\) meV.}
\label{Fig:3}
\end{figure}

\noindent
\textbf{\textit{Valley-dependent conductance}} - The analytical derivation of valley dependent conductance (Eq. \ref{eq:7}) allows us to analyze the whole conductance profile of the system for varying applied potential barrier height. The Fig. \ref{Fig:2} shows the valley-dependent conductance profile of the system. The angular dependence of the transmission profiles of each valley at \(  V_0=V_{K(K') } \) are shown in (c-f). The total conductance shown in (a) is calculated by considering continuous dispersion along all three directions. We observed that the highest valley polarization \({P_{K(K')}}\)  occurs in the conductance gap region (i.e. \(40< {V_0} < 60\) meV). However, this would be of little physical utility since the conductance of both valleys is negligibly low. Therefore, it is useful to introduce an effective valley polarization factor \({\beta _{K(K')}} = {P_{K(K')}} \times {G_{K(K')}}\) in order to evaluate the optimal conditions for high valley polarization and high conductance simultaneously. In Fig. \ref{Fig:2} (b) and (d), we consider the  \({\beta _{K(K')}}\)  factor for an exemplary device configuration where \({B_0} = 2\) T, \({E_F} = 50\) meV, \({w_y}/{v_F} = \chi \)0.4 and found that the effective valley polarization can be switched and its magnitude controlled by tuning the potential barrier height \({V_0}\). Numerically, the high polarization of \(K'\) can be selected by setting \({V_0} \approx \)110 meV which is denoted by \({V_{K'}}\) in Fig. \ref{Fig:2} (d). For the  \textit{K} valley, high polarization occurs at \( {V_0} \approx -20\) meV. At these voltage values \({V_K}\) and \({V_{K'}}\), the inequivalent transmission profile can be clearly seen by analyzing the valley resolved angular dependence of transmission probability \(T_{\phi ,\gamma }^{K(K')}\), as shown in Figs. 2 (c) to (f). The applied potential \({V_K}\) combined with the effect of magnetic barrier suppresses the electron transmission for both valleys. However, the electrons around \(K'\) possess much more restricted allowed transmission angle range compared to the electrons around \(K\)(see Fig. \ref{Fig:2} (c) and (d)). On the other hand, the applied voltage \({V_{K'}}\) allows limited angle range for transmission at \(K\), while barrier allows transmission almost the whole incident angle range at \(K'\), as shown in Fig. \ref{Fig:2} (e) and (f). This result indicates that the obtained valley polarization originates from electro-optical mechanisms, i.e., the refractions and reflections at the barrier interface. 

\noindent
\textbf{\textit{Fermi surface oriantation}} - To investigate the origins of the valley dependent tunneling shown above, we focus on the Fermi surface structure of the system at the barrier interface. The conservation of energy, momentum and chirality determines the ballistic electron transmission at the barrier interface. Therefore, one can predict the allowed incident angles that lead transmission by analyzing the matching of the Fermi surface and spin alignment at the barrier interface. In Fig. \ref{Fig:3}, the solid-lined (dashed) circle represents the Fermi surface in the presence (absence) of applied electrical potential \({V_0} = {V_{K'}} = 110\) meV, which is the barrier height for high \({\beta _{K'}}\). Due to the opposite direction tilt of the Weyl nodes at \(K\) and \(K'\), the Fermi surfaces shift to opposite transverse directions as shown in Fig. \ref{Fig:3} (a) and (b). At \(K\), the wave vector along transmission direction within the barrier, \({q_x}\) becomes imaginary for the electrons whose incident angle is between \(\pi /2\) and \({\phi _a}\). On the other hand, this forbidden angle range is between \( - {\phi _a}\) and \( - \pi /2\) for the electrons around \(K'\). This means that the electrons experience opposite direction deflections at the barrier interface according to their valley index. The magnetic barrier also causes a very similar shift of the Fermi surfaces. However, the effect of the magnetic barrier is valley-independent, which creates the same direction transverse shift for both valleys by means of transverse Lorentz displacement. As shown in Fig. \ref{Fig:3} (c) and (d), this additional one-way shift constricts the allowed angle range at \(K\), while it widens the allowed range at \(K'\). That remarkably shows that the barrier selectively reflects the incident electrons according to their respective valley index, causing the valley-polarized tunneling at the barrier interface. Note that the allowed range shown by shaded angles in Fig. \ref{Fig:3} does not mean that all the incident angles in this range lead transmission since the tunneling at the barrier interface also requires the matching of the spin or pseudo-spin alignment. As can be seen in Fig. \ref{Fig:2} (e) and (f), tunneling transmission occurs between Weyl nodes with the same chirality in the case \({V_0} = {V_K}\), causing the perfect matching of spin alignment; thus the transmission profile is only determined by the conservation of transverse momentum. However, Klein tunneling (chiral tunneling) transmission occurs in the case of \({V_0} = {V_{K'}}\) since \({V_{K'}} > {E_F}\), which causes the different pattern of perfect transmission angles due to the different spin alignment of electron and hole states, as seen in Fig. \ref{Fig:2} (e) and (f).

\begin{figure}[H]
\includegraphics[width=89mm]{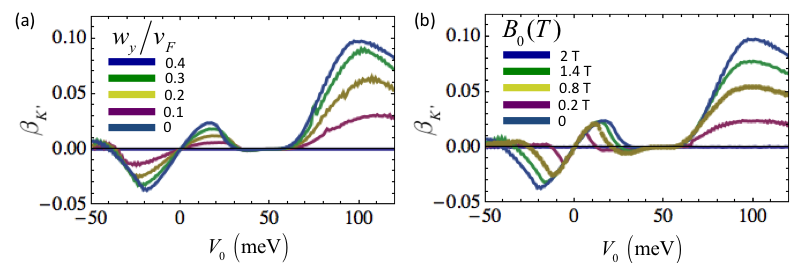}
\caption{\label{fig:epsart} The effect of tilt strength and applied magnetic barrier on the polarization of \(K'\). (a) shows the \({\beta _{K'}}\) for different tilt strength, where \({B_0}_{(z)} = 2\) T, (b) shows the \({\beta _{K'}}\) for different strength of applied magnetic barrier, where \({w_y}/{v_F} =\chi \)0.4. The Fermi energy \({E_F} = 50\) meV, and barrier length \(L = 900\) nm for all configurations.}
\label{Fig:4}
\end{figure}

To analyze the mechanisms that generate the valley-polarized transmission, we focus on the tilt strength of the Weyl nodes and applied magnetic barrier strength. As shown in Fig. \ref{Fig:4} (a) the polarization of valleys highly depends on the tilt strength that generates the valley-dependent transverse shift at the barrier interface. Note that this anomalous momentum shift is caused by the combined effect of the electrical potential and tilted band structure. The presence of only one of these factors would not lift the valley degeneracy. This combined effect lifts the valley degeneracy and separates the electrons in the two valleys in angle-space. However, the contribution of the both valleys to the overall conduction would be still same, leading to zero valley polarization in conductance. We require the use of magnetic barrier to cause a valley-selective electro-magnetic effect at the barrier interface. Thus, the strength of the valley-polarization of the conductance will be dependent on the magnetic field strength as well, as shown in Fig. \ref{Fig:4} (b).

\noindent
\textbf{\textit{Realistic magnetic barrier}} - The magnetic barrier in our proposed system would only affect the electron momentum by means of fringe fields induced by the magnetization of the FM layer, and not via the induced exchange proximity effect. In the schematic diagram of the proposed system shown in Fig. \ref{Fig:1}, there is a dielectric layer placed between the Weyl semimetal and the FM layer. Hence the overlap between the wave functions of the FM layer and the Weyl semimetal which gives rise to the exchange proximity effect is effectively suppressed if the separation distance (dielectric thickness) is more than a few nanometers. However, the fringe fields induced by the magnetized FM layer would be effective along the device thickness. Here, we will delve deeper into the physics of the magnetic barrier structure and provide a detailed quantitative analysis based on a more realistic model.
To calculate a more realistic magnetic barrier, we have considered the following magnetic field profile, which is consistent with the experimental results \cite{RNR2}:

\begin{figure*}[t]
\includegraphics[width=150mm]{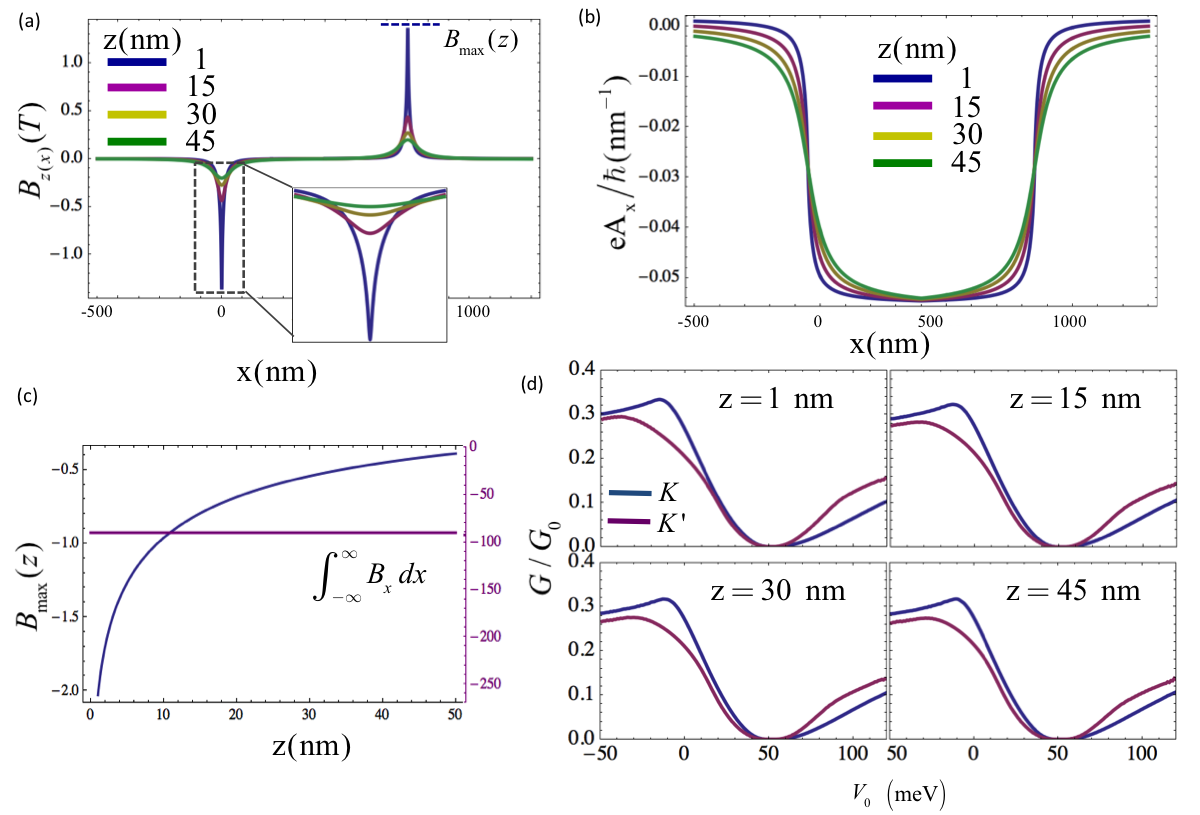}
\caption{\label{fig:epsart}  Realistic magnetic barrier configuration. (a) shows magnetic field profile with two asymmetric peaks, induced at the edges of the FM layer shown in Fig 1. Different colors represent the field profile at different points in the \textit{z}-direction. (c) shows the shift of the \(k_y\) due to the gauge potential \(A_x\) induced by the magnetic fields shown in (a). The maximum magnetic field strength \(B_{max}(z)\) shown in (a) reduces with increasing distance along \textit{z}, of which characteristic calculated for continuous distance in (c), while the integration of the magnetic field over \textit{x} is constant as shown at the right axis in (c). Since the maximum height of the \(A_x\) does not depend on the field profile, the valley resolved conductance shows very similar profiles at varying depth of the proposed device, as shown in (d).}
\label{Fig:5}
\end{figure*}

\begin{equation}
\
B_z(x)=\frac{\mu _0 M_s}{4}\ln (\frac{x^2+z^2}{(z+D)^2+x^2}).
\
\label{Eq:8}
\end{equation}

\begin{figure*}[t]
\includegraphics[width=150mm]{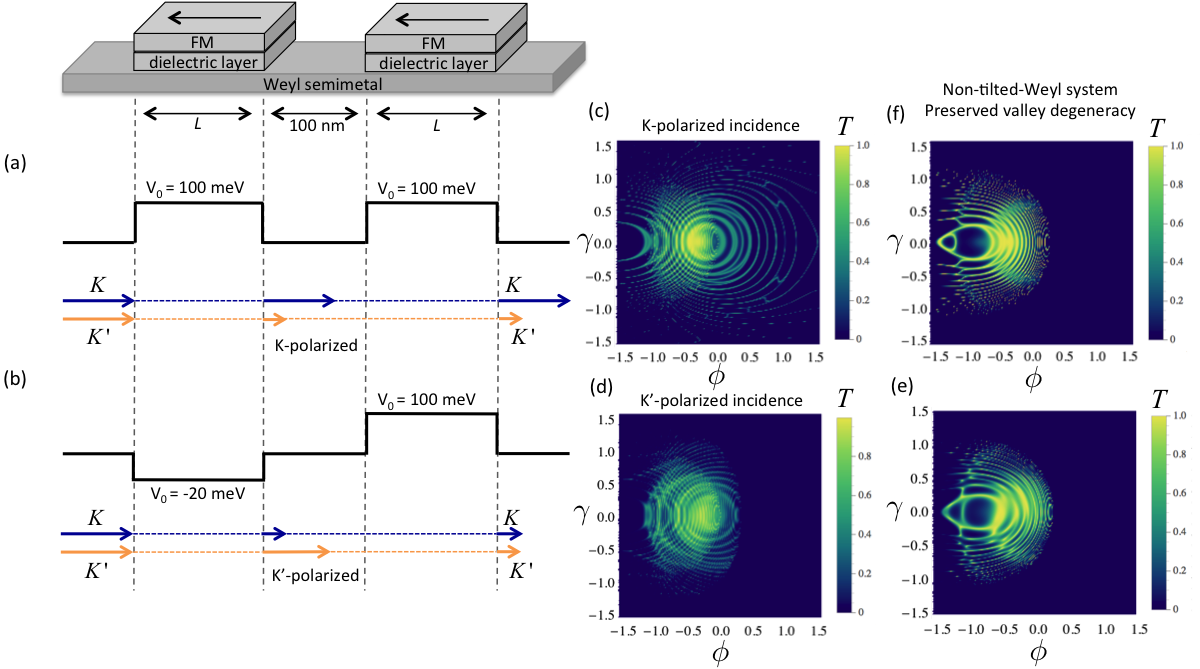}
\caption{\label{fig:epsart} Detection scheme for valley polarization, which is comprised of serially connected valley filters. Each one is identical to the scheme shown in Fig. \ref{Fig:1} (a). The first barrier is used as polarizer, while the second barrier is used to detect generated valley polarization. (a) and (b) shows two different configurations of the first barrier, i.e., set to the polarization of \(K\) and \(K'\) respectively. (c) and (d) show the angular dependence of transmission probability of the system illustrated in (a) and (b) respectively. (e) and (f) show the transmission probability for the valley-degenerate system \((w_y=0)\) in the case of the configurations shown in (a) and (b) respectively. The Fermi energy is 50 meV, barrier length \(L = 900\) nm, magnetic field  \(B_0=2 \) T for all configurations.}
\label{Fig:6}
\end{figure*}

In the above, \(  M_s \) is the saturation magnetization of FM layer, \textit{z} is the thickness of the dielectric layer between FM layer and Weyl semimetal, \textit{D} is the thickness of the FM layer.

Based on the above equation, the FM layer on top of the Weyl semimetal induces two asymmetric spike-like magnetic fields at the edges of the top FM layer. The magnetic field strength at the peak reduces with increasing distance along \textit{z}-direction, as shown in Fig. \ref{Fig:5} (a), where the different colors represents different values of \textit{z}. In addition to the change in the maximum value of the magnetic field strength, the field profile also spreads out over a wider extent in \textit{x}. We note that the \( \int B_x dx\) is constant at all depth \textit{z} even though the peak value \(B_{max}(z)\) reduces along \textit{z} (see Fig. \ref{Fig:5} (c)). Note that the maximum height of the \(A_x\) does not depend on the \(B_{max}(z)\) but the integral \( \int B_{(x)} dx\). The proposed valley filtering method depends primarily on the Fermi surface overlap, and hence the maximum height of the \(A_x\) rather than the specific profile of  \(A_x\) over different depth \(z\). Thus, we envisage that the variation of the magnetic field strength along \textit{z} would not have a strong influence on the valley dependent conductance of the system. To verify this, we perform numerical simulation by applying the realistic magnetic barrier profiles as shown in Fig. \ref{Fig:5} (a) and (b) and obtain the conductance profile of the system by considering different depths along \textit{z}. The transmission probability is obtained by dividing the whole device into short segments where the magnetic gauge potential is spatially varying along \textit{x}, and applying the transfer matrix method (Ref. \cite{CY7Datta}). The conductance is calculated by using Eq. \ref{eq:7}. The results shown in Fig. \ref{Fig:5} (d) reveal that the valley dependent conductance is highly robust against variation in the magnetic field profile along both \textit{x} and \textit{z}. For comparison, the saturation magnetization of the FM layer is set to the value that induces the same gauge potential height with that considered in Fig. \ref{Fig:2}. The conductance profiles in (d) are quite consistent with the result [Fig. \ref{Fig:2} (a)] obtained by considering a square shape gauge potential as described above, and which can be calculated analytically by the wave-function matching method. 
\\

\noindent
\textbf{\textit{Tuning the valley-polarization by gated potential barrier}} - The one of the advantages of the proposed valley filter function is that its operation is not restricted to some specific parameter range. Theoretically, the application of electrical barrier in tilted band structures causes a transverse momentum shift at the Fermi level and this shift is generally valley dependent. Similarly, any magnetic field along a particular direction, regardless of its strength or profile would break the angular symmetry of transmission by means of transverse Lorentz displacement. Thus, as can be seen from the Fig. \ref{Fig:2} (a) and (c), non-zero valley polarization can be achieved at any arbitrary applied gate voltage. However, the strength of the polarization can indeed be optimized by tuning the value of the gate voltage. As depicted in Fig. \ref{Fig:1}(d), the \(  K \) and \(  K' \) valley polarization can be optimized by setting the gate voltage at \(  V_K \) and \(  V_{K'} \). These optimal voltage values are a dependent on the applied magnetic field, which is tunable, as well as the intrinsic Fermi level which can be set by, e.g., by alkali metal doping \cite{RNR13}. We believe that given the flexible parameter configuration for the operation of the proposed system, the generated valley polarization could be readily observed.
Based on the parameter set demonstrated in Fig. \ref{Fig:2}, we require a change of the electrical potential of approximately 100 meV in order to switch the polarization between the two valleys. Such a change of potential can be achieved in a realistic system. For instance, let us consider gating by means of solid electrolytes \cite{RNR12} which is experimentally conducted in \(  Cd_3As_2 \). The electron and hole densities have been tuned to values on the order of \(10^{12} cm^{-2}  \), where a change of Fermi level in excees 100 meV, i.e., between 143-254 meV, was achieved under an applied gate voltage change of 0 to 12 V, based on experimentally observed Shubnikov-de Haas oscillations \cite{RNR12}. A similar experiment also confirmed that the carrier density rises up to values on the order of \( 10^{12} cm^{-2}  \)  by application of gate voltage in \(  WTe_2 \) \cite{RNR14}.

\noindent
\textbf{\textit{Effect of the finite device thickness on the valley polarization}} - 

Thus far, we assumed the condition where the finite thickness \(d_x\) is much larger than the Fermi wavelength \(\lambda_F\). This allows us to use the continuum treatment without considering the quantization of \textit{k}. However, having a small finite thickness may be a crucial factor if one requires the use of gated potential barriers due to the short range screening effect. Therefore, in this part, we consider the case where the system has a finite thickness along one of the directions and calculate the valley dependent conductance.

In this calculation, we apply the generic model (Eq. \ref{Eq:9}) that describes two Weyl nodes located at  \(k_x=k_y=0\),  \(k_z=\pm\Delta k_z/2 \) , where  \(\Delta k_z\) is the  \textit{k}-space distance between the two nodes.

\begin{equation}
\
H=\varepsilon_0 +\left(\begin{array}{cc} M(\bold{k}) & Ak_+\\  Ak_- & -M(\bold{k}) \end{array}\right)
\
\label{Eq:9}
\end{equation}

\begin{figure}[h]
\includegraphics[width=70mm]{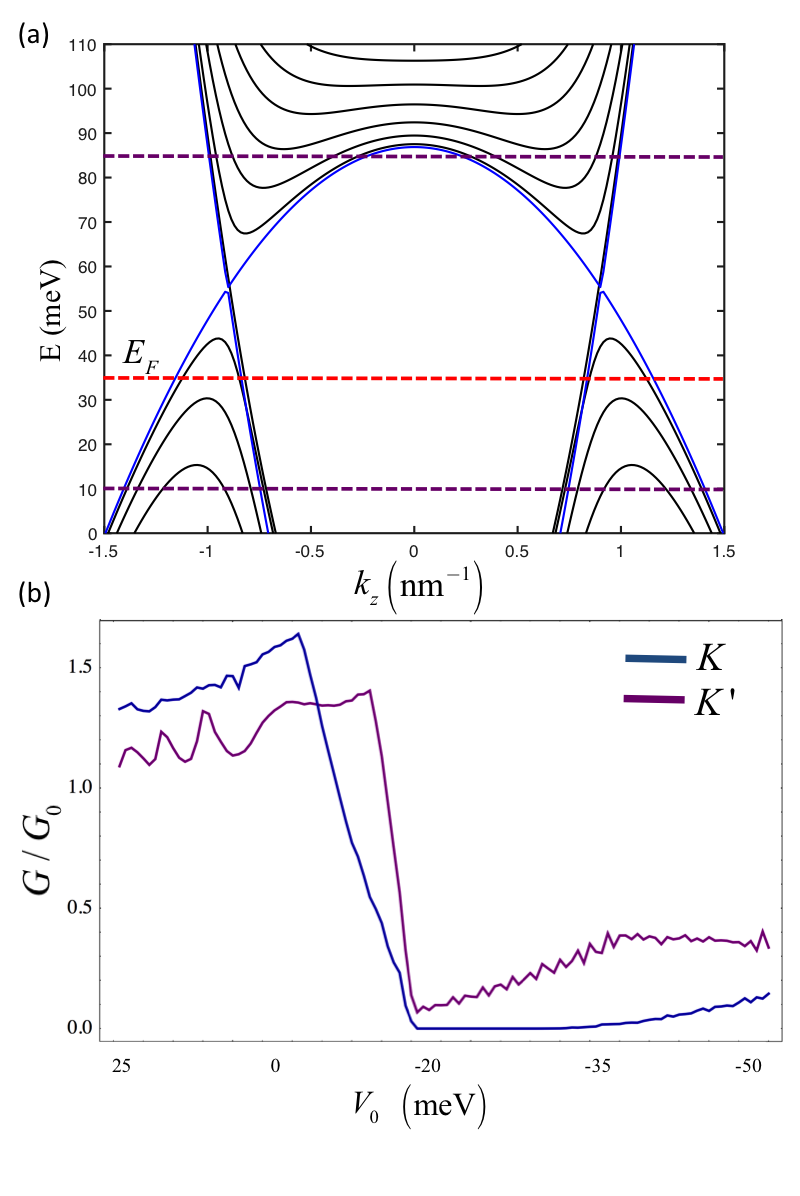}
\caption{\label{fig:epsart}(a) The energy dispersion at \(k_y=0\), where the \(E_F=35\) meV which is represented by the red dashed line. Two Weyl nodes appear at \(k_z \simeq \pm0.9 \textrm{ nm}^{-1}\). In our model, the application of a top gate voltage generates electrostatic barrier height \((V_0 )\) in the central region. The conductance of the system shown in (b) is calculated for \(-50 \textrm{ meV}<V_0<25 \textrm{ meV}\), which covers the applied potential range between the purple dashed lines. These lines indicate the levels in the central region where the Fermi level coincides with, in the case of  \(V_0=25 \textrm{ meV}\) and \(V_0=-50 \textrm{ meV}\). The material parameters used in the calculations are \(C_0=0, C_1=67.538 \textrm{ meV} \textrm{ nm}^2\), \(C_2=-84.008 \textrm{ meV} \textrm{ nm}^2\), \(M_0=-86.86 \textrm{ meV}\), \(M_1=-106.424 \textrm{ meV} \textrm{ nm}^2\), \(M_2=-103.610 \textrm{ meV} \textrm{ nm}^2\), \(A=245.98 \textrm{ meV}\).}
\label{Fig:7}
\end{figure}

In the above, \(\varepsilon_0=C_0+C_1 k_z^2+C_2 (k_x^2+k_y^2 )\), \(M(k)=-M_0+M_1 k_z^2+M_2 (k_x^2+k_y^2 )\) and \(k_\pm=k_x\pm ik_y\). Since the parameters, i.e., \(A\), \(C_i\), and \(M_i\) are material-specific; we take the example of \( \textrm{Na}_3\textrm{Bi}_2\), except \(C_1\) which is modified to match the tilt velocity of our system for comparison with the analytical result presented.

The finite thickness \(d_x\) along the \textit{x}-direction leads to sub-band dependent mass due to the quantization of \(k_x\) such that \( \left\langle k_x^2\right\rangle_n \approx \frac{n \pi}{d_z}  \), where \(n\)=1,2,{\ldots}.  In order to obtain the eigenspectrum, the Schrödinger equation can be expanded in the basis of the \(\langle x  \vert  \psi \rangle=\sqrt{2/d_x }\cos(n\pi x/d_x) \) infinite quantum well eigenstates and diagonalizing the resultant matrix (a more detailed explanations has been presented in Refs. \cite{RNR1,CY4,CY5}). The energy dispersion and quantized sub-bands of a system consisting of two Weyl nodes are shown in Fig. \ref{Fig:7} (a). The band structure also holds in the Dirac semimetal case with degenerate spins without coupling\cite{CY1,CY2}. Therefore, the proposed valley filter is not restricted to Weyl semimetals and it can be also realized in Dirac semimetals with tilted energy dispersion. As shown in Fig. \ref{Fig:7} (a), the Weyl nodes exhibits tilted characteristic along the \textit{z}-direction, which results in elliptical Fermi surfaces. When the energy shifts, the mismatch occurs between Fermi surfaces in the source and the drain regions, as shown in the illustration in Fig. \ref{Fig:3}. Based on the requirements of the proposed valley filter, we chose the transmission direction along \textit{y}-direction, where the \(k_z\) is a good quantum number. The conductance profile of a system depicted in Fig. \ref{Fig:1}, which consists of a 900-nm central barrier region sandwiched between the semi-infinite source and drain regions, is calculated numerically by matching of the wave functions at the barrier interfaces. A magnetic barrier is applied to the central region having the same profile as that  described in Fig. \ref{Fig:1}. 
	As plotted in Fig. \ref{Fig:7} (b), there is an imbalance in the conductance profiles of \(K\) and \(K'\) due to the combined effect of the electrical potential and the magnetic field. In the region between \(V_0\simeq-5\) meV and \(V_0\simeq25\) meV, the conductance of \textit{K} is higher than \(K'\). This region largely coincides with the homogeneous junction case (p-p*-p). On the other hand, the conductance of \(K'\) is higher than \(K\) in the remaining range, which mostly coincides with heterogeneous junction case (p-n-p). This result is in close agreement with the results obtained earlier by the continuum treatment (Fig. \ref{Fig:2}) where the finite thickness \(d_x\) is much larger than the Fermi wavelength \(\lambda_F\), and hence the quantization of \(k\) is ignored.

\noindent
\textbf{\textit{Detection of the generated valley polarization}} - Valleytronic applications have been reported in graphene-based structures, and various detection schemes have been presented for detecting the valley current in such systems. It has been experimentally shown in graphene systems with broken inversion symmetry, that valley-polarized current may cause inverse valley Hall effect that can be measured as a transverse voltage drop \cite{RNR6,RNR7,RNR8}. Alternatively, superconducting contacts can be used to detect valley polarized current based on Andreev reflection in systems where the valleys are related by time-reversal symmetry \cite{RNR9}. These schemes may be adapted for Weyl semimetal materials where the inversion symmetry is broken, while time-reversal symmetry is preserved. 

Besides, the valley-polarized current can also be detected in our present scheme by including a second barrier in series with the first (similar to that shown in Ref. \cite{RNR10}). Using the wave functions, and energy dispersion derived previously, the tunneling transmission can be calculated by the matching of the wave functions in the five regions at their respective barrier interfaces. The transmission probability is calculated numerically by the transfer matrix method. The (second) barrier that is used for detection must be set to a configuration where only one valley is allowed for transmission (e.g. 100 meV in Fig. \ref{Fig:2} (c)). To implement the detector scheme, valley polarized transmission is generated at the first barrier by tuning the electrical potential of the proposed valley polarizer. Fig. \ref{Fig:6} (a) and (b) shows two different configurations where the valley polarizer is set to the polarization of \(  K \) and \(  K' \) respectively. This can be achieved by tuning the electrical potential to -20 meV and 100 meV, which are the voltage values correspond to high polarization of \(  K \) and \(  K' \) respectively, based on the conductance profiles shown in Fig. \ref{Fig:2}. In the Fig. \ref{Fig:6}, the detector is set to measure \(  K \)-polarization. The total transmission of the system clearly shows larger transmission when the incident current is \(  K \)-polarized [see Fig. \ref{Fig:6} (c) and (d)]. To prove that the transmission difference originates from the valley polarization, we test the same barrier structures [Fig. \ref{Fig:6} (a), and (b)] in the case of \( (w_y=0)\), where the two valleys are degenerate in \( k\)-space. For the given parameters, two different configurations produce almost the same transmission profile. \\

\noindent
\textbf{\textit{Existance of multiple pair of Weyl nodes}} - Since the number of Weyl nodes, their respective chirality and tilt direction can vary according to the host materials, we have focused on the simplest Weyl semimetal case where the only one pair exists related by inversion symmetry. The presented scheme is applicable to this form of Weyl semimetals which have been widely investigated \cite{RNR15,RNF1,RNF2,RNF3,RNF4}. To generalize our valley polarization study to the case of multiple pairs of Weyl nodes, we have to consider the fact that it is only the component of the tilt vector that is perpendicular to the transmission direction that is responsible for the valley-dependent angular separation at the barrier interfaces.
 
Based on this, we analyze the possible cases of Weyl semimetals depending on their chirality and tilt direction. The Weyl semimetal phase requires either broken inversion symmetry (a) or time reversal symmetry (b). One can basically analyze these cases as follows:

(a) If one Weyl node exists at \textbf{\(K\)} whose energy dispersion is described by  \(  H=\hbar(v_Fk.\sigma+w.k) \), another Weyl node must appear at \textbf{\(-K \)} with opposite chirality due to the inversion symmetry, and its dispersion will be described by  \(  H=\hbar(-v_Fk.\sigma-w.k) \). Similarly, in the case of multiple pairs, as a result of inversion symmetry, Weyl nodes with different chiralities would possess tilt vectors in the opposite directions. In this case, the proposed filter serves a dual function, in that it would not only polarize the electrons according to their valley index, but also according to their respective chiralities. 

(b) If one Weyl node is located at  \textbf{\(K\)} described by  \(  H=\hbar(v_Fk.\sigma+w.k) \), the other Weyl node must appear at  \textbf{\(-K\)}  related by time reversal symmetry, which may be described by   \(  H=\hbar(v_Fk.\sigma-w.k) \). In this case, there exists at least another pair of Weyl nodes in the Brillouin zone. However, in this case, further information such as the position of the Weyl nodes in \textit{k}-space, their respective energy levels, must be accounted for in calculating the net valley polarization. These properties of the Weyl nodes is dependent on the crystal symmetries and material structure of the Weyl semimetal.

Therefore, we take the example of HgTe class materials to demonstrate the possible configuration that generates valley dependent conductance based on our approach. Recent works have shown that the HgTe class of materials host four pairs of Weyl nodes which are type-I or type-II according to the strength of the applied compressive strain\cite{CY12}. By tuning the strain, the Weyl cones can be slightly tilted in the case of type-I, which is the case considered here.

\begin{figure}[h]
\includegraphics[width=70mm]{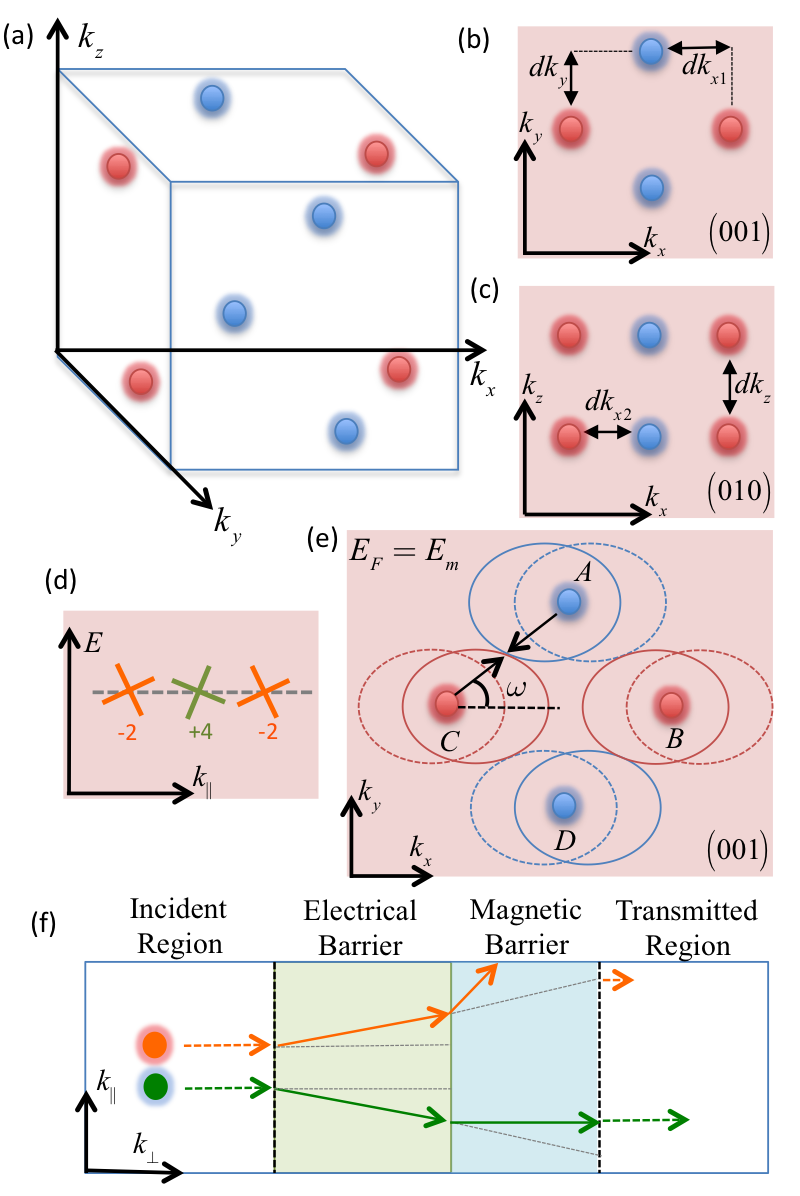}
\caption{\label{fig:epsart} Weyl node configuration of HgTe under compressive strain that generate four pairs of type-I Weyl fermions, where the chirality of the Weyl nodes are represented by the red and blue colors. (b) and (c) shows the reflection of the Weyl nodes on the (001) and (010) surface Brillouin zones. (d) schematically illustrates the two groups of Weyl nodes classified according to their tilt directions. Each of the green and orange groups includes four Weyl nodes. (e) shows the Fermi surface of the Weyl fermions, where the dashed-lined circle represents the Fermi surface of the same cone under electrical gate potential. The examplary configuration in (e) indicates the upper limit of the Fermi energy where there is no overlap between Fermi surfaces of different valleys. The proposed transport mechanisms is illustrated in (f), which effectively polarize, and filter the desired group of valleys [i.e., orange or green shown in (d)] by means of electrical and magnetic barriers. For simplicity, the magnetic and electrical barriers are illustrated sequantially., In the proposed model they operate simultaneously in the same barrier region.}
\label{Fig:8}
\end{figure}

Based on the Weyl node locations in HgTe shown in Fig. \ref{Fig:8} (a), the following points are important considerations for the possible experimental realization of the proposed valley filter.

i) As the proposed approach is demonstrated by assuming ballistic tunneling transport, the Fermi surfaces around the Weyl nodes with different chirality must not be overlapping along \textit{k} in the transmission direction. Otherwise, not only the momentum, but also the spin orientation in the vicinity of the valleys would influence the tunneling transmission. Although this may not totally suppress the valley dependent conductance, it may reduce the tunneling conductance due to the mismatch of the spin orientation. In this respect, HgTe is a suitable candidate for observing our proposed effect since there is no overlap between the valleys with different chiralities if one chooses the transmission direction to be along \( k_x\) or \( k_y\) or \( k_z\) as shown in Fig. \ref{Fig:8} (a-c).

ii) The Weyl nodes with tilt vector pointing in opposite direction must not be overlapping along the transmission direction since the proposed method polarizes the valleys according to the tilt direction. HgTe would satisfy this under specific strain configuration that results in the ideal Weyl node distribution shown in Fig. \ref{Fig:8}, as discussed in Ref. \cite{CY12}.

iii) The \textit{k}-space distance of the Weyl nodes is a crucial factor due to the inter-valley scattering, which may reduce the valley polarization. However, the current Weyl semimetal candidates possess large enough \textit{k}-space separation that effectively decouple the Fermi surfaces, and hence suppress the transmission between the valleys with opposite direction tilt. We analyze this in detail in a later section.

Consequently, based on the above analysis, the Weyl nodes can be characterized into two groups according to the direction of their tilt vector, and are represented by the green and orange cones in Fig. \ref{Fig:8}. (d). Due to the opposite tilt direction of the nodes in the transverse direction, the electrons close to different valleys would experience a deflection in opposite directions according to their respective groups, as indicated by the green or orange arrows in Fig. \ref{Fig:8} (f). The transverse Lorentz displacement due to the magnetic barrier additionally deflects the electrons along the transverse direction, which results in valley dependent conductance across the barrier. In Fig. \ref{Fig:8} (f), for simplicity, the electric and magnetic barriers are illustrated sequentially. In actual fact, they are effective simultaneously in the same (central) region in the proposed model.

Based on the crystal structure of the materials, the two valley groups shown in Fig. \ref{Fig:8} (d) and (f) may be formed such that valleys in a group would share both the same tilt direction and chirality. In this case, our proposed model would yield electrons which are polarized according to the valley index, as well as chirality (i.e., they exhibit both valley and chirality polarization). This case occurs where tilt at a node is correlated with its chirality (e.g., the simplest Weyl semimetal phase with a single pair of nodes related by the inversion symmetry). We could also have the other scenario where valleys in each group only share the same tilt direction but not the chirality direction. In this case, our model would yield electrons which are only valley polarized. The detection scheme proposed in the manuscript (which is sensitive to the valley index) would be suitable for both cases. In the example of HgTe shown in Fig. 8, each orange and green group contains both chiralities (shown by blue and red). Thus the generated current would be polarized according to valley index, but not the chirality.

Consequently, even in such the case of multiple pairs of Weyl nodes, we would in general expect finite valley polarization to occur. Based on our proposed model, valleys that possess different tilt direction can be angularly separated with a single electrostatic barrier. Note that the tilt vectors of Weyl nodes does not need be exactly opposite to one another to allow valley polarization to occur. Valley polarization can be achieved as long as the transmission and tilt directions are not exactly parallel to each other. As a result, by choosing the transmission direction and the direction of the magnetic field according to the number and the position of the Weyl nodes, one can still generate valley-polarized transmission with multiple Weyl node pairs based on our proposed method.

\noindent
\textbf{\textit{k-space separation of the valleys}} - As mentioned earlier, the robustness of the Weyl semimetal case strongly depends on the \textit{k}-space distance of the Weyl nodes, i.e., the length of the Fermi arc. The clustering of Weyl points close in \textit{k}-space would negatively affect the valley polarization in the proposed scheme due to the inter-valley transmission of Weyl fermions. However, in terms of ballistic transport, intervalley transmission requires Fermi surface (FS) overlap due to the conservation of the transverse momentum. To investigate the effect of the Weyl node separation on the presented results, we consider the limiting case of the FSs where the overlap starts to occur between the propagating states of the different Weyl nodes (assuming low energy approximation). Since the Weyl node orientation in \textit{k}-space depends on the crystal structure of the material, we focus on the specific HgTe example shown in Fig. \ref{Fig:8}. The four Weyl nodes located at \( k_z=\pm k_z^*\) are perfectly overlapping along the choosen transmission direction \(k_z\), which is a desirable condition as the momentum and the spin orientations are perfectly matched. The FS orientation is shown in Fig. \ref{Fig:8} (e), where the solid-lined circles represents the FSs out of the barrier region, and dashed-lined circles represent the FSs within the barrier region. In HgTe, the Weyl points are located at \((\pm k_x^*, 0, \pm k_z^*)\) and \((0, \pm k_y^*, \pm k_z^*)\) where \(k_x^*=k_y^*=k^*\). Due the choice of transmission direction along \(k_z\), transmission between different Weyl nodes would be excluded if there is no overlap of the FSs on the \(k_x\)-\(k_y\) plane. The limiting condition is shown in Fig. \ref{Fig:8} (e) where the FSs just touch each other with no overlap. Note that there is also relatively small momentum shift \textit{dk} induced by the magnetic barrier (not visible in Fig. \ref{Fig:8} (e)), which shift the dashed-lined FSs further. To derive the analytical expression that describe the limitting condition, we consider without loss of generality on the specific Weyl node pair A and C [see Fig. \ref{Fig:8} (e)]. The touching point between the FSs of A and C can be found as \(k_\omega=(E_m-V_0)/(\hbar(v_F+w_x \cos(\omega)  ) )\), where \(\omega\) is the angle shown in Fig. \ref{Fig:8} (e) which is given by \(\omega\)= \( \tan^{-1}((k^*/(k^*+dk)) \). And, the \textit{k}-space distance between A and C is \( \sqrt{(k^*+dk)^2 +(k^* )^2 }\), which is also equal to \(2k_{\omega} \). Thus, one can find the upper limit of the Fermi energy that avoids an overlap between FSs, which is given by

\begin{equation}
\
E_m=\frac{V_0}{2} + \frac{\hbar}{2} ( v_F+\frac{w_x}{\xi} )\sqrt{(k^*+dk)^2 +(k^* )^2 }
\
\label{Eq:10}
\end{equation}
\begin{equation}
\
\xi=\sqrt{1+\frac{(k^* )^2}{(k^*+dk )^2}}
\
\label{Eq:11}
\end{equation}

In HgTe with particular strain configuration, \(k^*=0.073 \textrm{nm}^{-1}\)(see Ref. \cite{CY12}) which gives that \(E_m\simeq22.5\) meV in the case of the example configuration (i.e., consistent with the presented demonstration of the valley polarization) where \(V_0=0 \) meV, \(dk\simeq-0.055 \textrm{ nm}^{-1}\), \(w_y=0.4v_F\). We note that the inter-node separation can be much larger in other Weyl semimetals, e.g., in \(\textrm{Ta}_3\textrm{S}_2\) it is as large as \(~1.5 \textrm{ nm}^{-1}\) (see Ref. \cite{CY14}), which implies that the Fermi energy can be as large as \(\simeq 496\) meV without incurring overlap for the same configuration of \textit{dk}, \(w_x\), and \(V_0\). These Fermi energy values are already higher or comparable to what we assume in our calculations. Furthermore, we note that the proposed scheme does not impose a tight restriction on the Fermi energy. The system can operate over a wide range of Fermi energy by optimizing other parameters such as the magnetic field.

\section{\label{sec:level1}CONCLUSION}

In summary, we presented a new type of valley filter approach that is compatible with all the systems containing tilted band structure. The presented effect originates from the coupling of the applied potential gradient and valley dependent tilt of energy dispersion around the Weyl points. Further investigation of this valley polarization effect may pave the way for novel applications in electron optics of Weyl semimetals. Valley-polarized tunneling applications have been proposed in many systems under the influence of uniaxial strain, such as in graphene and silicene, but modulation of the applied strain is not so straightforward. Unlike previous works, our analysis shows that the possibility of controlling valley-polarized tunneling by means of electrical bias in Weyl systems possessing tilted band structure, which is more readily achieved in practice. The parameter used in this work can be further optimized to achieve higher valley polarization.

\section{\label{sec:level1}ACKNOWLEDGEMENT}

The authors would like to acknowledge the MOE Tier II grant MOE2013-T2-2-125 (NUS Grant No. R-263-000-B10-112), and the National Research Foundation of Singapore under the CRP Program ``Next Generation Spin Torque Memories: From Fundamental Physics to Applications'' NRF-CRP9-2013-01 for financial support.

\appendix
\renewcommand\thefigure{\thesection.\arabic{figure}}    
\section{}
\setcounter{figure}{0}    

\renewcommand{\theequation}{\thesection.\arabic{equation}}
\setcounter{equation}{0}    

\begin{figure}[H]
\centering
\includegraphics[width=60mm]{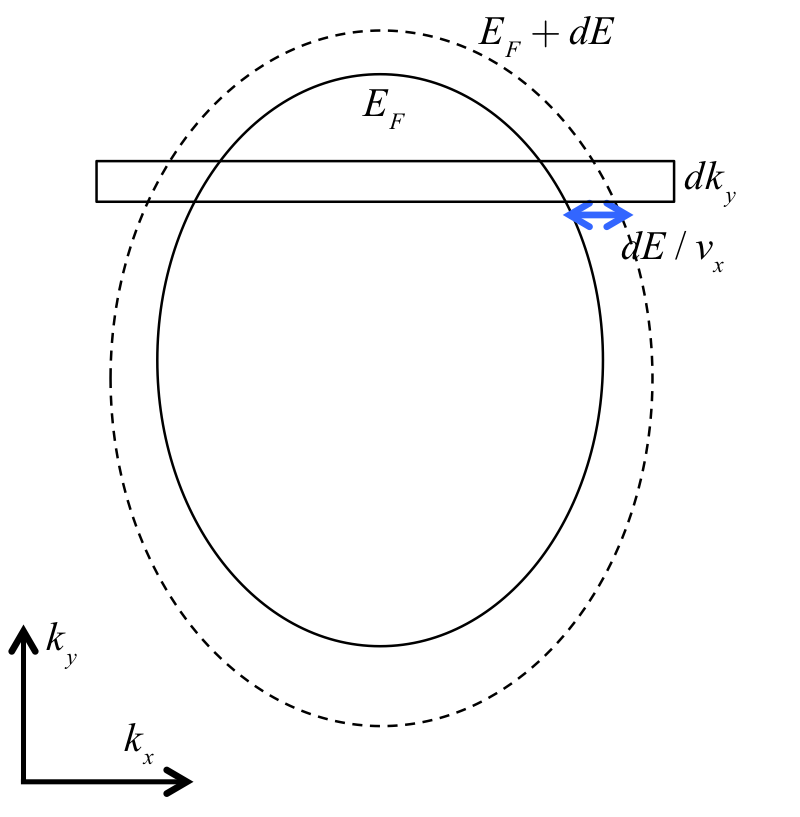}
\caption{\label{fig:A1} The solid ellipse represents a Fermi surface; and the dotted ellipse shows  \(E_F+dE\)}
\label{Fig:A1}
\end{figure}

In systems with two or three dimensions, the conservation of the transverse momentum can be treated as a parameter with a given value. To calculate the transmission, one can sweep through the range of the transverse momentum covered by the source Fermi surface and at each value of the transverse momentum. This is illustrated by the diagram (Fig. \ref{Fig:A1}), which shows a two-dimensional system on the \textit{xy} plane where the interface is along the \textit{x} direction and \(k_y\) is conserved. 

\begin{figure}[H]
\includegraphics[width=89mm]{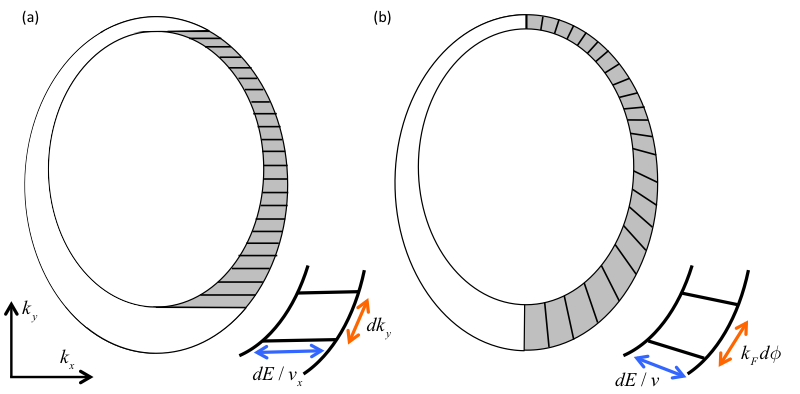}
\caption{\label{fig:A1} (a) shows the integration over \(dk_y\) where the dimensions of the infinitesimal element are \(dk_y\) and \(\frac{dE}{v_x}\). (b) shows the integration over \(\phi\) where the dimensions of the infinitesimal element are \(k_F (\phi)d\phi\) and  \(\frac{dE}{v}.\)}
\label{Fig:A2}
\end{figure}

For simplicity we assume \(\hbar=1\) in the following derivations. At each given value of \(k_y\), the \textit{k}-space area between the \(E=E_F\) and \(E=E_F+dE\) surface is \(  \frac{1}{v_x }  dE dk_y\). In the Landauer-B{\"u}ttiker formalism \(dE=eV_B\) where the \(V_B\) is the applied bias between the source and drain. The number of states per unit real-space area per \(dk_y\) is then \(\frac{1}{(2\pi)^2 }   \frac{1}{v_x}  eV_B\). Considering the transmission at each value of \(k_y \) as \(T(k_y )\) then the drain current for each state is \( \frac{e}{L}  T(k_y ) v_x^d (k_y )\), and the total current would then be given by

\begin{equation}\label{eq:A1}
  I/(eV_B)=\int dk_y \frac{eW}{(2\pi)^2 } \frac{v_x^d(k_y)}{v_x^s(dk_y) } T(k_y) 
\end{equation}

\noindent
where \(W\) is the width of the system. The above integration is over the Equal Energy Contours (EECs) between the Fermi surface, and the EEC at \(E_F+eV_B\) where \(v_x\) is positive. This is the shaded area in panel (a) of Fig. \ref{Fig:A2} where the infinitesimal area element consists of horizontal strips. The panel (b) in Fig. \ref{Fig:A2} shows the integration over the incident angle \(d\phi\). In the limit that dE goes to zero, the shaded areas in (a) and (b) are identical. The infinitesimal area element in (b) is \(k_F (\phi)d\phi dE/v)\).

In our system, the coordinates(\(\phi\),\(\gamma\)) are defined as

\begin{equation}\label{eq:A2}
\begin{array}{ccl}
x=k_F\cos(\gamma)\cos(\phi)\\
y=k_F\cos(\gamma)\sin(\phi)\\
z=k_F\sin(\gamma) 
\end{array}
\end{equation}

Thus, the 3-dimensional analog of the Eq. \ref{eq:A1} would then be

\begin{equation}\label{eq:A3}
  I/(eV_B)=\int\int_{FS} dk_y dk_z \frac{eA}{(2\pi)^3 } \frac{v_x^d(k_y, k_z)}{v_x^s(k_y, k_z) } T(k_y) 
\end{equation}

\noindent
where the subscript FS stresses the fact that the \(k_y\) and \(k_z\) values are to be chosen from points lying on the Fermi surface only. The above equation can be re-written as follows in terms of \(\phi\) and \(\gamma\).

\begin{equation}\label{eq:A4}
  I/(eV_B)=\int\int dS_{FS} \frac{eA}{(2\pi)^3 } \frac{v_x^d}{v} T(\phi, \gamma) 
\end{equation}

In the above, note that the density of the states is related to \(1/v\) instead of \(1/v_x\) . The infinitesimal area element per unit variation of \(\phi\) and \(\gamma\) is

\begin{equation}\label{eq:A5}
d{S_{FS}} = \frac{{E_F^2\cos \gamma \sqrt {v_F^2 + (\chi w_y)^2 + 2{v_F}{\chi w_y}\cos \gamma \sin \phi } }}{{{\hbar ^2}{{\left( {{v_F} + {\chi w_y}\cos \gamma \sin \phi } \right)}^3}}}d\phi d\gamma.
\end{equation}

The velocity operators are defined as \(\widehat{v}_x=v_F \widehat{\sigma}_x\), \(\widehat{v}_y=v_F \widehat{\sigma}_y + \chi w_y \widehat{I}_\sigma \), \(\widehat{v}_z=v_F \widehat{\sigma}_z\). Their corresponding expectation values of the velocity operators are given by

\begin{equation}\label{eq:A6}
\begin{array}{ccl}
\langle \widehat{v}_x \rangle=v_F\cos(\gamma)\cos(\phi)\\
\langle \widehat{v}_y \rangle=v_F\cos(\gamma)\sin(\phi)+\chi w_y\\
\langle \widehat{v}_z \rangle=v_F\sin(\gamma) \\
v= \sqrt{\langle \widehat{v}_x \rangle^2+\langle \widehat{v}_y \rangle^2+\langle \widehat{v}_z \rangle^2}\\

\end{array}
\end{equation}

Assuming identical source and drain and restoring the \(\hbar\), the conductance can be found as

\begin{equation}\label{eq:A7}
G=G_0\int\int d\phi d\gamma   \frac{\cos^2(\gamma) \cos(\phi) }{(1+ \frac{\chi w_y}{v_F} \cos(\gamma) \cos(\phi) )^3  }  T(\phi, \gamma)
\end{equation}

\noindent
where  \( G_0= \frac{e^2 E_F^2 A}{(2\pi)^3 \hbar^3 v_F^2} \)  is the quantum conductance, and the integral being dimensionless.

\bibliographystyle{unsrt}

\end{document}